\documentclass{sig-alternate-2013}

\newif\iflongver
\longvertrue 

\newfont{\mycrnotice}{ptmr8t at 7pt}
\newfont{\myconfname}{ptmri8t at 7pt}
\let\crnotice\mycrnotice%
\let\confname\myconfname%

\permission{Permission to make digital or hard copies of all or part of this work for personal or classroom use is granted without fee provided that copies are not made or distributed for profit or commercial advantage and that copies bear this notice and the full citation on the first page. Copyrights for components of this work owned by others than ACM must be honored. Abstracting with credit is permitted. To copy otherwise, or republish, to post on servers or to redistribute to lists, requires prior specific permission and/or a fee. Request permissions from permissions@acm.org.}
\conferenceinfo{WPES'14,}{November 3, 2014, Scottsdale, Arizona, USA.}
\copyrightetc{Copyright 2014 ACM \the\acmcopyr}
\crdata{978-1-4503-2948-7/14/11\ ...\$15.00.\\
http://dx.doi.org/10.1145/2665943.2665946}

\usepackage{amsmath}
\usepackage{amssymb}
\usepackage{theorem}
\usepackage{amsfonts}
\usepackage{color}
\usepackage{graphicx}
\usepackage{picture}
\usepackage{graphics}
\usepackage{booktabs}
\usepackage{xcolor}
\usepackage{caption}
\usepackage[margin=5pt]{subcaption}
\usepackage{tikz}

\iflongver
    \makeatletter
    \def\@copyrightspace{\relax}
    \makeatother
\fi

\newcommand{\set}[1]{\mathcal{#1}}        

\newcommand{\event}[1]{\langle #1 \rangle}      

\newcommand{\pr}[1]{\mathbb{P}\mathrm{r}\{#1\}}

\newcommand{\ex}[1]{\mathbb{E}\{#1\}}

\newcommand{\deltap}{d_p}      
\newcommand{\deltaq}{d_q}      
\newcommand{\dqmax}{Q_{loss}^{\max}} 

\definecolor{gray}{RGB}{190,190,190}
\definecolor{sgreen}{RGB}{46,139,187}

\newcommand{\miniuser}{\includegraphics[scale=0.15]{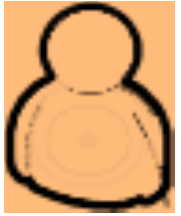}}

\newcommand{\etal}{\emph{et~al.}}

\begin{document}

\title{Prolonging the Hide-and-Seek Game:\\ Optimal Trajectory Privacy for Location-Based Services\iflongver \\{\small (extended version)}\fi}
\numberofauthors{1}
\author{
\alignauthor
George Theodorakopoulos$^1$, Reza Shokri$^2$, Carmela Troncoso$^3$, \\Jean-Pierre Hubaux$^4$ and Jean-Yves Le Boudec$^4$ \\ \medskip
       \affaddr{$^1$Cardiff University, UK, $^2$ETH Zurich, Switzerland, $^3$Gradiant, Spain, $^4$EPFL, Lausanne, Switzerland}, \\
       \email{\footnotesize $^1$g.theodorakopoulos@cs.cardiff.ac.uk, $^2$reza.shokri@inf.ethz.ch, $^3$ctroncoso@gradiant.org, $^4$firstname.lastname@epfl.ch}
}

\clubpenalty = 10000
\widowpenalty = 10000

\maketitle

\begin{abstract}
Human mobility is highly predictable. Individuals tend to only visit a few locations with high frequency, and to move among them in a certain sequence reflecting their habits and daily routine. This \emph{predictability} has to be taken into account in the design of location privacy preserving mechanisms (LPPMs) in order to effectively protect users when they continuously expose their position to location-based services (LBSs).
In this paper, we describe a method for creating LPPMs that are customized for a user's mobility profile taking into account privacy and quality of service requirements. By construction, our LPPMs take into account the sequential correlation across the user's exposed locations, providing the maximum possible \emph{trajectory} privacy, i.e., privacy for the user's present location, as well as past and expected future locations. Moreover, our LPPMs are optimal against a \emph{strategic adversary}, i.e., an attacker that implements the strongest inference attack knowing both the LPPM operation and the user's mobility profile.
The optimality of the LPPMs in the context of trajectory privacy is a novel contribution, and it is achieved by formulating the LPPM design problem as a Bayesian Stackelberg game between the user and the adversary. An additional benefit of our formal approach is that the design parameters of the LPPM are chosen by the optimization algorithm.




\end{abstract}



\section{Introduction}
Location-Based Services (LBSs) provide users with valuable information about their surroundings such as traffic status (e.g., Beat the Traffic, or INRIX Traffic Maps, Routes \& Alerts), nearby points of interest (e.g., Google Maps), or friends' activities (e.g., Foursquare or Google Latitude). Despite this benefit, information about our current, future, or frequently-visited locations is highly sensitive, as it can be used to infer our habits, preferences, political and religious affiliations, as well as to endanger our physical security if it falls in the wrong hands.

Hence, the need arises to protect the location privacy of LBS' users, while maintaining the usability and quality of these services. The task at hand must account for the three following considerations:

\textbf{First} and foremost, as Shannon's maxim states, ``One ought to design systems under the assumption that the enemy will immediately gain full familiarity with them.'' In other words, the adversary will adapt his attack to the protection mechanism. This in turn shall lead to an updated mechanism, then to a novel attack, and so on ad infinitum. This is commonly known as the \emph{arms-race problem}. To cut the arms race short, in our approach the defender anticipates the adversary's reaction, and so the initial design is already robust against an informed adversary.

\textbf{Second}, in order to facilitate the deployment of a Location Privacy Preserving Mechanism (LPPM) users must be able to use it independently of other users' behavior, and without the permission or collaboration of a third party. In other words, a usable privacy-preserving mechanism must be user-centric, in the sense that decisions taken to protect privacy (e.g., hiding, perturbing, or faking the location) need to be made locally to the users. Our approach only requires users to perform a local look-up in a pre-computed table, and hence it can be easily integrated in mobile devices frequently used to access LBSs.

\textbf{Third}, protecting the user's current location is intricately bound with protecting her past and future locations. Different LBSs require the user location to be updated at different rates. Some require frequent updates (e.g., finding nearby friends, or obtaining live traffic information), while others can function perfectly well with just a single location (e.g., seeking nearby points of interest). The privacy protection offered by an LPPM is severely impacted by the frequency with which locations are revealed, since locations exposed in quick succession are highly correlated. When correlated locations are exposed, inferring the user's current location provides the adversary with tools to reduce the uncertainty on the user's immediate past or future whereabouts. Our LPPMs take location correlation into account to effectively protect a user's location privacy along her trajectory.

We propose a framework to design user-centric LPPMs that -- given a user's quality requirements, privacy requirements, and mobility profile -- can
(1) protect the privacy of past locations (i.e., the current obfuscation is chosen to be compatible with past ones),
(2) protect the privacy of future locations (i.e., the current obfuscation is chosen to be compatible with the likely next locations)
(3) protect the privacy of transitions between locations (i.e., even though two successive locations may not be individually sensitive, the act of going from the first to the second might be sensitive),
(4) protect locations that the user visits between two LBS accesses (i.e., locations that the user visits without issuing an LBS query from them).
To the best of our knowledge, this is the first work that addresses the two latter objectives as separate targets in need of protection. The two former objectives have already been addressed in the literature, but the protection techniques are non-optimal (e.g., they do not consider informed adversaries, or they do not operate in a user-centric manner).

Our LPPMs can find the optimal privacy protection in any given scenario, i.e., they provably achieve the best protection among all possible mechanisms against a strategic adversary with knowledge of mobility profiles and LPPM algorithm. In other words, an LPPM designed under our framework provides a level of privacy that constitutes an upper bound on the privacy that is achievable by any other defense. The proposed framework can handle a wide range of correlation levels without making any prior assumption on the user mobility and LBS access patterns: From cases where there is high correlation between exposed locations, such as location-based services that require continuous requests to the provider (e.g., navigating with Google maps), to cases in which there is complete independence among exposed locations, such as LBS that require sporadic location updates (e.g., checking-in in Foursquare).

The key technique used in our solution is a Bayesian Stackelberg game between the privacy defender and adversary, launched by the user every time she wishes to share her location with the LBS. The two main benefits of this game theory technique are the following. First, it can naturally express the objective of optimizing privacy under the double constraint of anticipating the adversary's attack and respecting the user's quality requirements. Second, it allows us to efficiently search in an infinite space of potential solutions, guaranteeing that the computed solution is optimal without testing each and every single one of them (computationally impossible) and without limiting ourselves to the heuristic (but undeniably creative) solutions that human ingenuity can concoct. We provide a general design method that can be instantiated for particular privacy objectives, periods of observation and patterns of exposure. For the sake of illustration, we also provide examples to specifically protect two plausible privacy objectives: protecting the most recent locations (including the current one) and protecting the current and future locations.

We run our solution on real users' trajectories to obtain optimal location obfuscation mechanisms that maximize privacy, defined as the adversary's error in estimating the user's true location. Our results show that our method is more effective at protecting location privacy than protection mechanisms that only consider the currently exposed locations. We also show that quality of service can be traded off for privacy, but the maximum privacy achievable is strongly dependent on users' behavior (i.e., on the predictability of their movements). 



\section{Related Work} \label{sec:relatedwork}

In this section, we qualitatively compare our approach to previous work on location privacy. This comparison focuses on schemes that provide trajectory privacy~\cite{ChowM11} rather than sporadic privacy~\cite{ShokriTTHB12,DamianiBS10,ChengZBP06}, i.e., privacy of locations exposed to the LBS independently of each other. A quantitative comparison with the latter is provided in Section~\ref{sec:evaluation}, where we show that our trajectory-aware approach outperforms sporadic privacy-preserving mechanisms when protecting trajectories.

A first class of trajectory-aware mechanisms in the literature are those that aim at protecting user privacy when trajectories are published in bulk. Protection is achieved by grouping trajectories of different users in a wide area to ensure that the aggregate trajectory can be ascribed to at least $k$ users~\cite{AbulBN08}; mixing the trajectories of $k$ users~\cite{NergizASG09}; eliminating some events from the published dataset~\cite{HohGXA10,TerrovitisM08}; or replacing locations with larger regions defined by a pre-defined grid~\cite{GidofalviHP07}. Along similar lines, some protection algorithms need access to the complete trajectory before protection can be applied~\cite{YouPL07}, or they delay the exposure of queries so as to gather additional information about subsequent user locations \cite{ghinita2009preventing,ardagna2012protecting}. In contrast, our approach decides in real time how to protect a given location that the user is about to expose.

Other trajectory-aware mechanisms assume the existence of a trusted third party (e.g., the cellular service provider)~\cite{PanMX09, Gao13}, or assume that nearby users are present and can be leveraged to achieve joint privacy protection~\cite{BeresfordS03,FreudigerSH09,HuangYMS06}. Both of these scenarios violate the user-centricity design requirement in this paper. \iflongver Not depending on other users is also the reason why $k$-anonymity does not apply in our case, as well as any other method that attempts to make a user indistinguishable from other users. \fi

In addition to addressing trajectory privacy in a user-centric and real-time manner, our major qualitative difference from prior work is that we address the problem formally. This has two main advantages. First, it allows to provide provably optimal privacy against any adversary describable by our model. Second, it allows to define fine-grained knobs for expressing the user's privacy and quality requirements. We achieve these properties by formalizing the privacy-preserving mechanism design problem as a Bayesian Stackelberg game, similarly to Shokri~\etal~\cite{ShokriTTHB12} who focused on sporadic exposures. We would like to emphasize that our approach is not an extension to Shokri~\etal, but uses the same methodology to achieve a different goal, i.e., to protect trajectory privacy instead of sporadically exposed locations.

The only other formal approach that we are aware of is Andr{\'e}s~\etal~\cite{AndresBCP12}, who extend the concept of differential privacy to location privacy, thus defining a new privacy metric: geo-indistinguishability. They also propose a mechanism to achieve it. Similarly to Shokri~\etal, Andr{\'e}s~\etal~focus on sporadic disclosures, stating that in the case of successive location disclosures, the geo-indistinguishability that their mechanism provides decreases linearly in the number of disclosed locations. However, this solution is not optimal and can be outperformed by \cite{ShokriTTHB12} in the case of assuming a prior. 
In Section~\ref{sec:evaluation}, we show that our algorithm compares favorably against that of Shokri~\etal~when protecting trajectory privacy, hence we can conclude that our algorithm also offers better protection than that of Andr{\'e}s~\etal. Chatzikokolakis~\etal~\cite{chatzikokolakis2014predictive} extend the geo-indistinguishability framework to mobility traces. However, the solution is not optimal. Bordenabe~\etal~\cite{bordenabe2014optimal} provide the solution for constructing optimal differential private obfuscation mechanisms. Shokri~\cite{shokri2014optimal} designs user-centric obfuscation mechanisms that are optimal both with respect to the differential privacy metric and the prior leakage from the user. These optimization problems, however, are not applied on trajectory privacy. Thus, to the best of our knowledge, our work is the first formal solution to optimize trajectory privacy.

\iflongver
    \section{Trajectory Privacy}\label{sec:lbs}

Users' movements are not isolated discrete events. Rather, users follow a path to arrive from one place to another, following a \emph{trajectory}. Along this trajectory, users may query a location-based service to obtain useful information concerning the surroundings or the arrival point. Queries may be launched continuously in time, or only at selected spots. The former reveals the whole trajectory to the service provider, while the latter hides some parts. However, even if not all points in the trajectory are exposed to the service provider, the correlation between consecutive positions implies that inferring just one of them reveals information about past and future ones. For instance, spatio-temporal constraints (e.g., maximum user velocity), or road configuration and direction, may reveal with high probability the route followed by a user between two successive location exposures. The higher the level of correlation, the more information is revealed.

The correlation between successively shared locations depends on two factors: randomness of user mobility patterns and LBS access frequency. The former relates to how predictable a user's future location is given her current location. The latter defines the rate at which the LBS provider can sample the user's trajectory.
These two factors act multiplicatively on correlation, but they have opposite effects. On the one hand, high randomness decreases correlation between successively exposed locations, since the current position contains less information about past and future events than when movements are deterministic. On the other hand, high LBS access frequency increases correlation, since regardless of the randomness of her movements the user has little time to move between two LBS accesses and exposed locations are nearer to each other than when access frequency is low.



\begin{figure}
        \centering
        \begin{subfigure}[t]{0.22\textwidth}
                \centering
                \includegraphics[width=\textwidth]{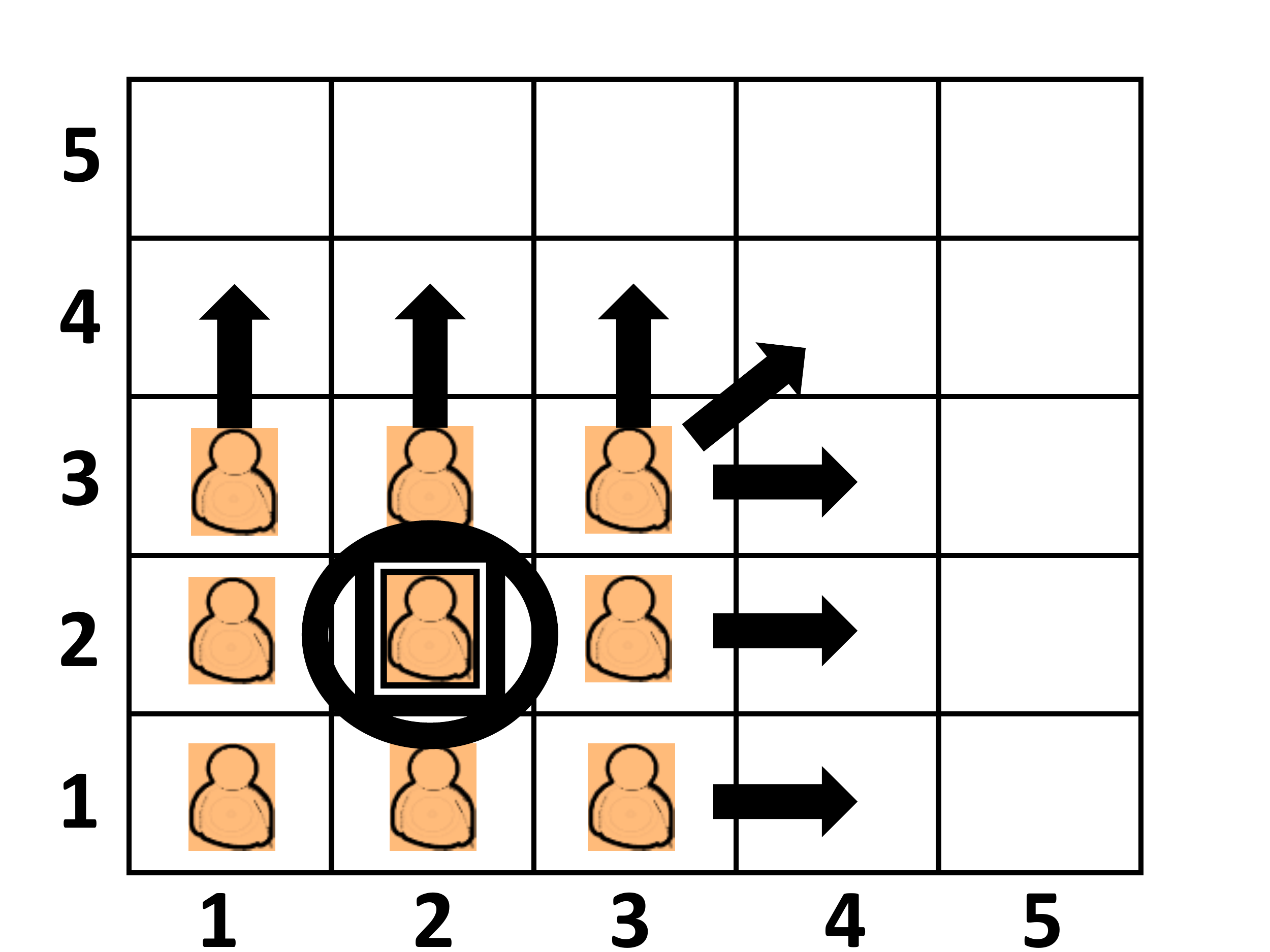}
                \caption{\textbf{Time $t-1$}: real location (2,2) ($\bigcirc$); exposed pseudolocation (2,2) ($\square$). Since the user moves only to adjacent locations, at time $t$ she will be in the bottom left 4x4 square.}
                \label{fig:fig1t-1}
        \end{subfigure}
        \begin{subfigure}[t]{0.22\textwidth}
                \centering
                \includegraphics[width=\textwidth]{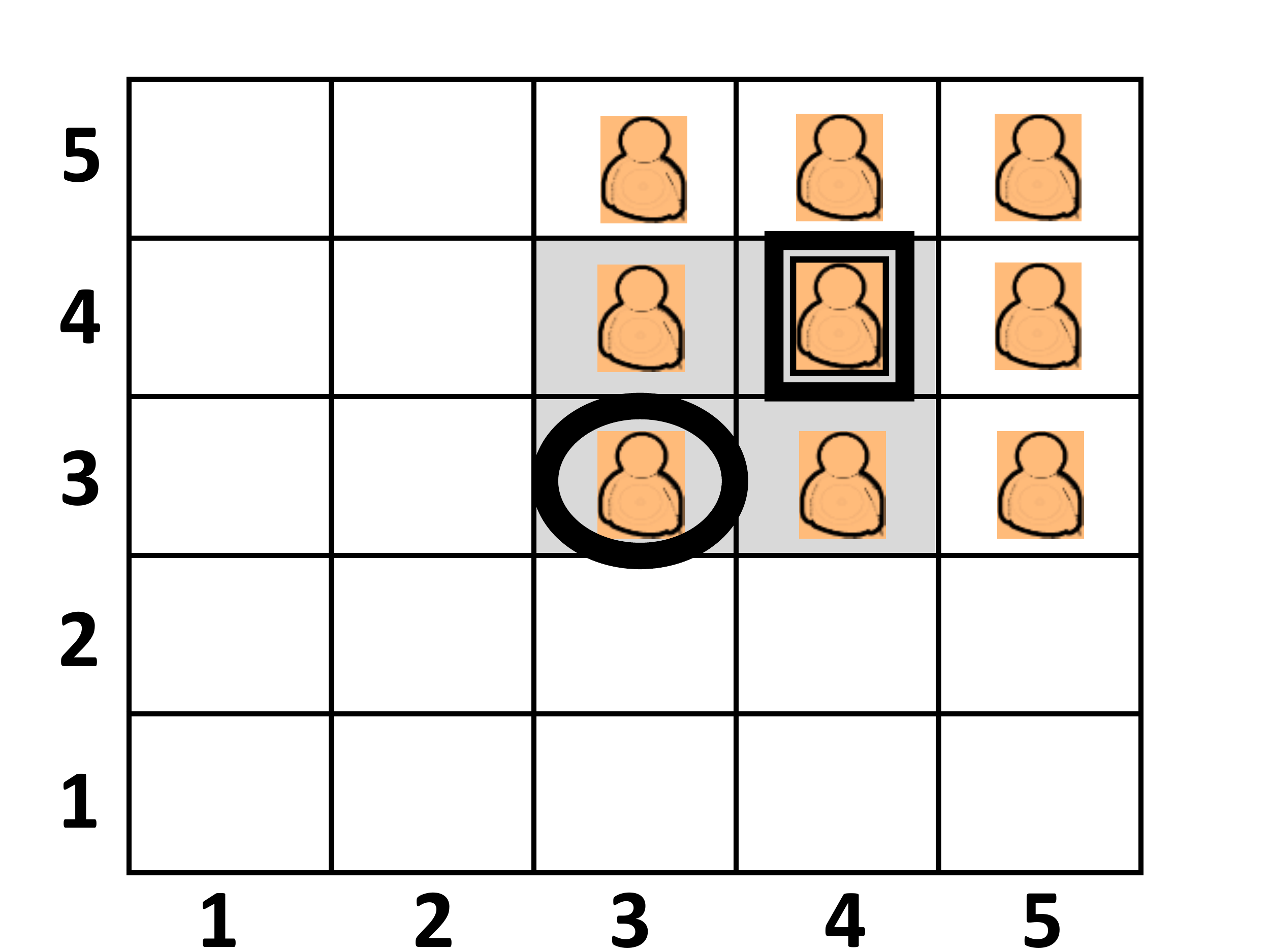}
                \caption{\textbf{Time $t$}: real location (3,3) ($\bigcirc$); exposed pseudolocation (4,4) ($\square$). The user can only be in the dark 2x2 square, rather than the 3x3 square (\miniuser).}
                \label{fig:fig1t2}
        \end{subfigure}
        \caption{Ignoring correlation when choosing exposed pseudolocations can compromise the user's current location.} \label{fig:example_correlation1}
\end{figure}

Considering correlation between exposed locations is of utmost importance when designing privacy protection mechanisms for location-based services in which the location has to be exposed frequently to the LBS provider. We now show, through a toy example, that ignoring the correlation between successive locations leaks information that allows the adversary to reduce her uncertainty about past and/or future locations of the user. Let us consider that a user moves around in a 5x5 grid, moving at most one location per time unit (see Figure~\ref{fig:example_correlation1}). This user accesses an LBS using an LPPM that, given a real location $(x,y)$, outputs a pseudolocation chosen arbitrarily from the 3x3 square centered on $(x,y)$, $\{(x+i,y+j),\, i=\{-1,0,1\}, j=\{-1,0,1\}\}$. This pseudolocation is then sent to the LBS. The LBS, which observes the exposed pseudolocations, tries to infer the user's movements using prior knowledge of the LPPM algorithm and of the user behavior.

First, assume that the user has accessed the LBS from location (2,2) at time $t-1$ and the LPPM reported pseudolocation (2,2), as shown in Figure~\ref{fig:fig1t-1}. The adversary can infer that the user could only have been in the bottom-left 3x3 square of locations -- these are the only locations from which the LPPM may output (2,2), represented by \miniuser. Moreover, since the user can move by at most one location per time unit, the adversary knows that at time $t$ she will be somewhere inside the bottom-left 4x4 square (see Figure~\ref{fig:fig1t-1}).

The user accesses the LBS again at time $t$ from location (3,3), reporting pseudolocation (4,4). Naively, one may think that the probability of the adversary correctly guessing her real location is 1/9 (a random location in the 3x3 square surrounding location (4,4)), similarly to time $t-1$. However, given the prior observation the adversary knows that at time $t$ the user can only be in the bottom-left 4x4 square. Intersecting this knowledge with her current observation the adversary can deduce that the user is in the darkened 2x2 square in Figure~\ref{fig:fig1t2}. Therefore, the probability of a correct guess is 1/4, more than twice as much than the naively expected 1/9. This example highlights that choosing pseudolocations \emph{disregarding correlation may reduce the privacy of the current location}.

Now consider the example in Figure~\ref{fig:example_correlation2}, where the LPPM reports (1,1) at time $t-1$ instead of reporting (2,2), and reports (4,4) at time $t$. In this case, the real locations at \emph{both} $t-1$ \emph{and} $t$ are completely compromised: The only two-step trajectory that is compatible with the successive exposure of pseudolocations (1,1) and (4,4) is that the user accessed the LBS from (2,2) followed by (3,3). Strikingly, the \emph{privacy of past locations was retroactively compromised}: it was safe until the pseudolocation at time $t$ was reported.

\begin{figure}[htbp]
\begin{center}
    \begin{subfigure}[t]{0.22\textwidth}
                \centering
                \includegraphics[width=\textwidth]{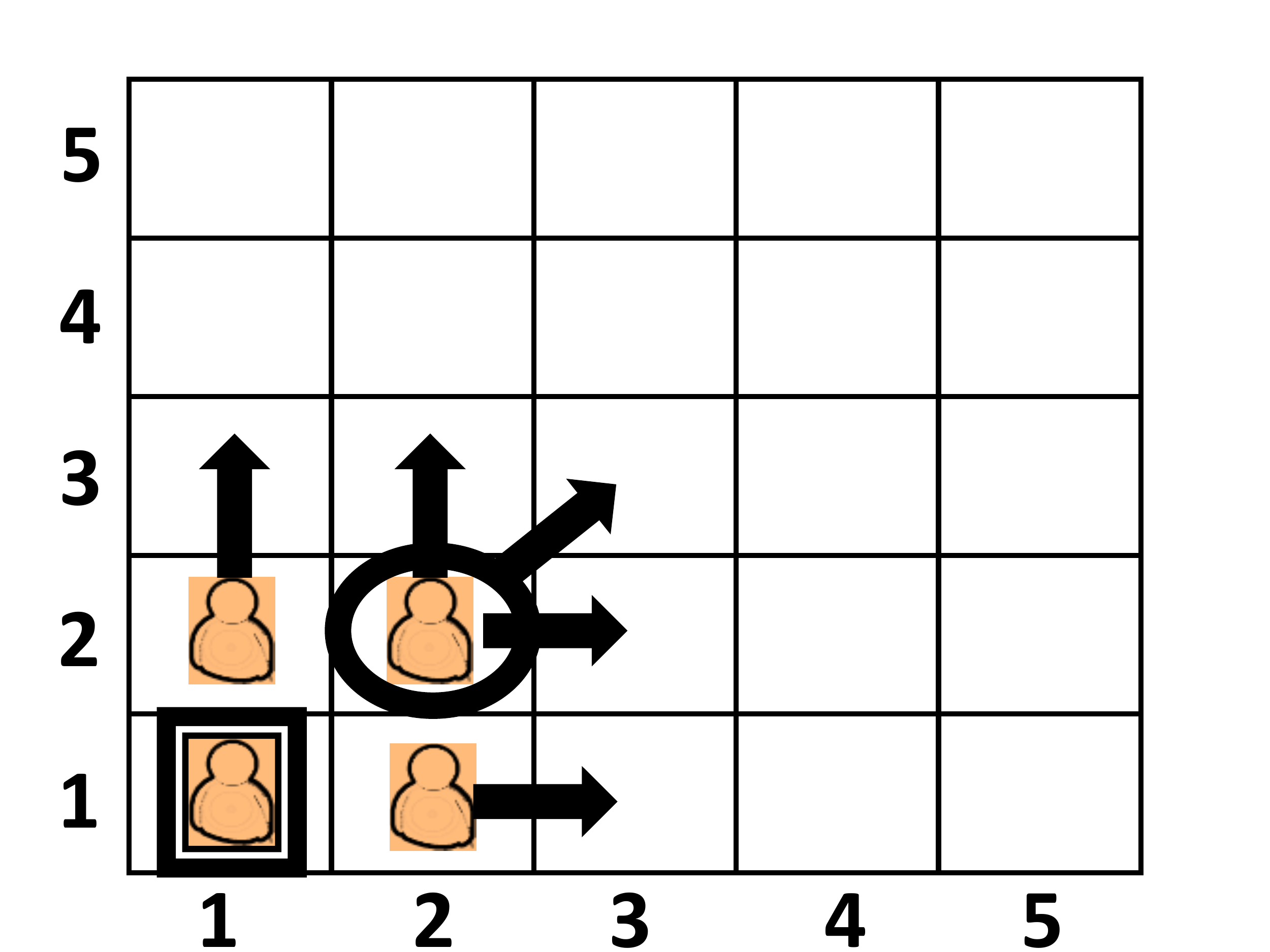}
                \caption{\textbf{Time $t-1$}: real location (2,2) ($\bigcirc$); exposed pseudolocation (1,1) ($\square$). Since the user moves only to adjacent locations, time $t$ she will be in the bottom left 3x3 square.}
                \label{fig:2fig1t-1}
        \end{subfigure}
        \begin{subfigure}[t]{0.22\textwidth}
                \centering
                \includegraphics[width=\textwidth]{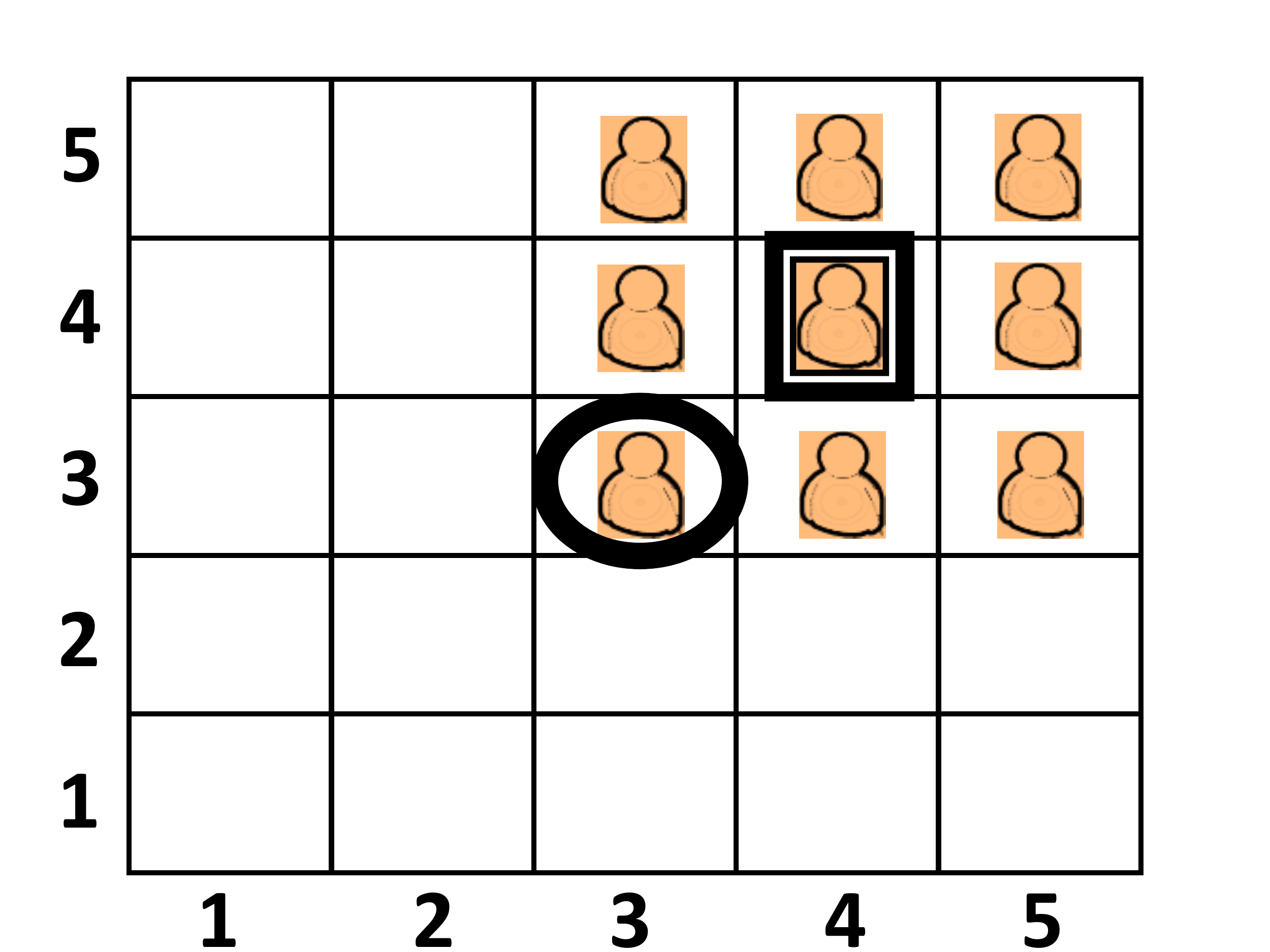}
                \caption{\textbf{Time $t$:}real location (3,3) ($\bigcirc$); exposed pseudolocation (4,4) ($\square$). The attacker can infer that at $t-1$ the user could only be at location (2,2), and at $t$ she can only be at location (3,3).}
                \label{fig:2fig1t2}
        \end{subfigure}
\end{center}
\caption{Naively choosing a pseudolocation may compromise not only the current but also the past location(s) of the user.}\label{fig:example_correlation2}
\end{figure}

Similarly, the pseudolocation chosen at time $t$ can affect future privacy. Consider a user at (2,2) at time $t$, who knows that she will go to (3,3) at $t+1$. If the LPPM chooses to expose pseudolocation (1,1) at time $t$, then at $t+1$ the user will not be able to expose (4,4) without revealing her real location at both $t$ and $t+1$. Hence, the choice at $t$ has an impact on the possible choices at $t+1$. In other words, \emph{future privacy may be proactively compromised by current choices}.

The conclusion from these examples, and the motivation for our design, is that LPPMs need to choose pseudolocations that are compatible both with previously exposed pseudolocations and with possible future movements. Compatibility means that, even if an adversary is aware of the LPPM's inner workings and of the user's general mobility pattern, he should not be able to infer the user's real locations from the exposed pseudolocations. Our design finds the optimal way to expose pseudolocations without restricting choice to a fixed pattern, e.g., to a uniform selection from a 3x3 square around the user. On the contrary, any pseudolocation can be chosen with any probability, aiming at maximizing privacy in the best possible way against an informed adversary.

\fi

\section{Problem statement}\label{sec:statement}

\paragraph*{User Mobility and LBS Access Pattern}

Consider a user moving within $M$ discrete locations $\set{R} = \{ r_1, r_2, \ldots, r_M \}$. The user's movements are represented as a discrete-time trajectory of locations at times $\set{T} = \{1, 2, \ldots\}$. An \emph{event} $\event{r,t}$ denotes that the user is at location $r \in \set{R}$ at time $t \in \set{T}$. Slightly abusing notation, the time-subscripted variable $r_t$ will denote the user's location at time $t$. Typical values from $\set{R}$ are $r, r_i, r_j$, whereas typical time-subscripted locations are $r_t, r_{t-1}, r_{t+1}$.

The mobility of the user is modeled probabilistically. In our implementation, we choose to model mobility as a first-order Markov chain on $\set{R}$, but this is not mandatory. Any other model is possible, as long as it allows us to compute probabilities of the user visiting various sequences of locations. We note that our contribution is not tied to the first-order Markov choice since our method can compute the optimal protection for any given mobility model.

As the user moves, she accesses the LBS at each time instant $t\in \set{T}$, i.e., from each location that she visits. We assume this for simplicity, and in the appendix we show that we can accommodate sparser LBS-access patterns. After all, the most interesting case is when LBS-access times are close enough to each other so that successively exposed locations \emph{are correlated}. If they are not, the problem becomes equivalent to sporadic location disclosure, studied in~\cite{ShokriTTHB12}.

\paragraph*{LPPM Functionality}

The user wants to protect her privacy from an adversary who observes the locations exposed to the LBS (so the adversary could be the LBS provider itself, an eavesdropper, other LBS users, etc.). Hence, she uses an LPPM that obfuscates her real locations before they are sent to the LBS. We model obfuscation as a replacement operation in which a fake location from a set $\set{R'}$ is sent to the LBS instead of the real location. We take $\set{R'}$ to be the same as $\set{R}$. We call these fake locations \emph{pseudolocations} or \emph{obfuscated locations} and denote them by $r'$. The corresponding events $\event{r',t}$ are termed \emph{pseudoevents}.

\begin{figure}
  \centering
  \includegraphics[width=\columnwidth]{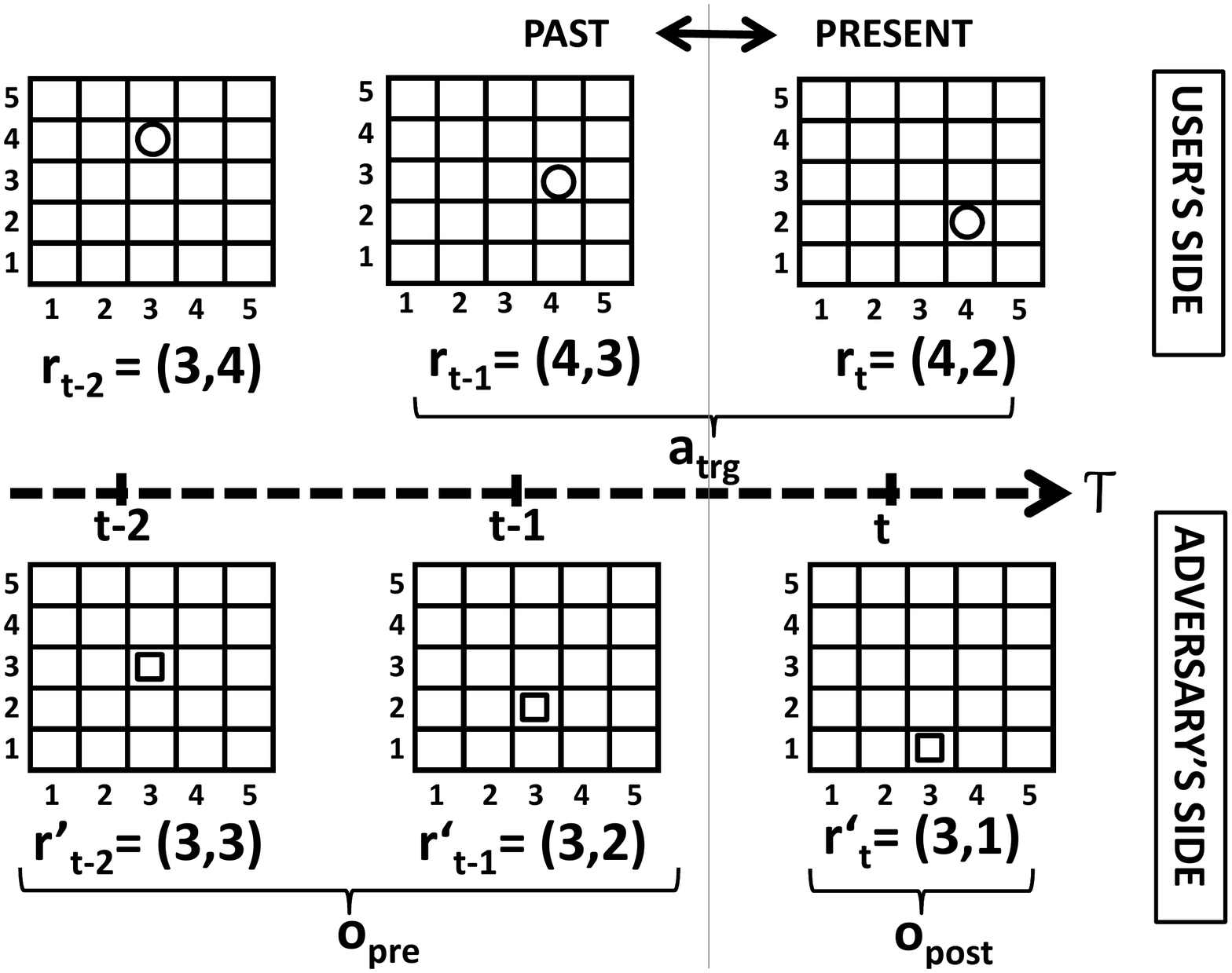}\\
  \caption{A user moves from location $(3,4)$ at time $t-2$, to (4,3) at $t-1$, to $(4,2)$ at current time $t$. The user wants to protect locations at times $t-1$ and $t$, and these are denoted by $a_\text{trg}$ (target events). At past times $t-2$ and $t-1$, the LPPM exposed pseudolocations $(3,3)$ and $(3,2)$ (denoted by $o_\text{pre}$), and to protect $a_\text{trg}$ the LPPM currently exposes location $(3,1)$ (denoted by $o_\text{post}$).} \label{fig:notation}
\end{figure}

The problem we tackle is the design of an LPPM algorithm $f(\cdot)$ that takes as input the real location (or locations) to be protected and the pseudolocations previously sent to the adversary, and then probabilistically selects the pseudolocation (or pseudolocations) to expose to the adversary.  The pseudolocation set is fixed, and it coincides with the location set. The adversary knows the probabilistic model that describes user mobility, and also knows the LPPM algorithm $f(\cdot)$. Notice here the self-reference: We design $f(\cdot)$ assuming an adversary who knows the $f(\cdot)$ that will be designed and can respond optimally to it.

We introduce the notation used throughout the paper with the example shown in Figure~\ref{fig:notation}. Assuming that the current time is $t$, the elements of the framework can be defined as follows:

\begin{itemize}
   \item $a_\text{trg}$ denotes the target events that the user wants to protect, or equivalently, the events that the adversary wants to infer. 
       In the example, the user wants to protect her location at times $t-1$ and $t$ and thus $a_\text{trg}=(r_{t-1}, r_t)=\{\event{(4,3), t-1},\event{(4,2), t}\}$.
   \item $o_\text{pre}$ is a subset of the pseudoevents that the LPPM created and sent to the LBS \emph{up to but before} the current time. These are the pseudoevents that matter for the estimation of $a_\text{trg}$: Typically, $o_\text{pre}$ would be a sequence of consecutive pseudoevents starting with a recent time instant (as old ones do not matter for estimating $a_\text{trg}$) and leading up to the current time. These are known both to the adversary and to the LPPM.
       In the example, the relevant pseudolocations were exposed at times $t-1$ and $t-2$ and thus $o_\text{pre}=(r'_{t-2}, r'_{t-1})=\{\event{(3,3), t-2},\event{(3,2), t-1}\}$.
   \item $o_\text{post}$ is the pseudolocation (or set of pseudolocations) that the LPPM produces to protect $a_\text{trg}$ and that will be sent to the LBS at current time. In the example, at current time $t$ the user exposes pseudolocation $(3,1)$ thus $o_\text{post}=(r'_t)=\{\event{(3,1), t}\}$.
   \item $f(o_\text{post} | a_\text{trg}, o_\text{pre})$ is the probability that the LPPM produces $o_\text{post}$, given its knowledge $o_\text{pre}$ and the locations $a_\text{trg}$ it is trying to protect. This function encodes the defensive mechanism. It can be viewed as a codebook that prescribes, for each value of $a_\text{trg}$ and $o_\text{pre}$, a randomization over the possible values of $o_\text{post}$.
\end{itemize}
Notice that $a_\text{trg}$ need not be the same length as $o_\text{post}$: In the example, the LPPM exposes the current (time $t$ only) pseudoevent, while aiming to protect the events of the current as well as the previous time instant ($t$ and $t-1$).

\paragraph*{Attacker Model and Privacy Metric}


%
%
%
%
%

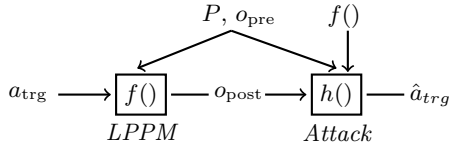
\begin{figure}[t]
\centering
\begin{tikzpicture} [auto=left,scale=0.5]
  \node[] (atrg) at (1,0) {$a_\text{trg}$};
  \node[draw,thick,label=below:{\em\footnotesize LPPM}] (lppm) at (4,0) {$f()$};
  \node[draw,thick,label=below:{\em\footnotesize Attack}] (adv) at (9.2,0) {$h()$};
  \draw[->][thick] (1.8,0) -- (3.2,0);
  \draw[thick] (4.8,0) -- (5.8,0);
  \node[] (opost) at (6.6,0) {$o_\text{post}$};
  \draw[->][thick] (7.3,0) -- (8.4,0);
  \draw[thick] (10,0) -- (11,0);
  \node[] (ahtrg) at (11.7,0) {$\hat{a}_{trg}$};
  %
  \node[] (opre) at (6.6,2.1) {$P$, $o_\text{pre}$};
  \draw[->][thick] (6.4,1.7) -- (3.9,0.7);
  \draw[->][thick] (6.4,1.7) -- (9.2,0.7); 
  \node[] (f) at (9.4,2.1) {$f()$};
  \draw[->][thick] (9.5,1.7) -- (9.5,0.6); 
\end{tikzpicture}
  \caption{Information available to the LPPM and the adversary: The LPPM wants to protect location(s) $a_\text{trg}$ by producing appropriate pseudolocation(s) $o_\text{post}$. The adversary observes the output $o_\text{post}$ of the LPPM and, using his knowledge of the LPPM function $f$, estimates $a_\text{trg}$; the adversary's estimate is $\hat{a}_\text{trg}$. The prior knowledge of the adversary and of the LPPM consists of the transition matrix $P$ and the pseudolocations $o_\text{pre}$ that have been produced in the past.}\label{fig:infoflow}
\end{figure}

In short, privacy is quantified as the adversary's error in estimating the user's true location(s) $a_\text{trg}$.
Figure~\ref{fig:infoflow} illustrates the information flow of events and pseudoevents to the LPPM and to the adversary. The detailed notation is as follows:

\begin{itemize}
   \item $\psi(a_\text{trg} | o_\text{pre})$ is the adversary's prior probability distribution on the inference target $a_\text{trg}$, given his prior knowledge $o_\text{pre}$. It encodes what the adversary can deduce about $a_\text{trg}$ \emph{before} observing the LPPM's current output $o_\text{post}$.
   \item $\hat{a}_\text{trg}$ denotes the adversary's estimate of $a_\text{trg}$. 
       Similarly to $a_\text{trg}$, it can be seen as a time-indexed vector whose elements belong to the set $\set{R}$ of locations.
   \item $h(\hat{a}_\text{trg} | o_\text{pre}, o_\text{post} )$ is the probability that the adversary estimates $\hat{a}_\text{trg}$ to be the true value of $a_\text{trg}$, given his knowledge of prior pseudolocations $o_\text{pre}$ and given the pseudolocation(s) $o_\text{post}$ exposed at current time $t$. Note that, by definition, $h(.)$ can contain multiple observed exposed pseudolocations that happen across any time period (e.g., multiple days).
   \item $\deltap(\hat{a}_\text{trg}, a_\text{trg}) \geq 0$ is the privacy gain when the adversary's estimate is $\hat{a}_\text{trg}$ and the true value of the inference target is $a_\text{trg}$. It is zero only if $\hat{a}_\text{trg} = a_\text{trg}$. The value of $\deltap$ for each pair of locations needs to be provided by the user in order to reflect the sensitivity of the user with respect to different location. The user needs to provide low values if she is sensitive towards a particular location $a_\text{trg}$. We treat $\deltap$ as an input to our framework. 
 \end{itemize}

The privacy that an LPPM $f(\cdot)$ achieves against an adversary implementing attack $h(.)$ is then the expected value of $\deltap(\hat{a}_\text{trg}, a_\text{trg})$, given prior observations $o_\text{pre}$:

\begin{align}\label{eq:privacygain}
&Privacy(\psi, f, h, \deltap; o_\text{pre}) = \ex{\deltap(\hat{a}_\text{trg}, a_\text{trg}) | o_\text{pre}}
\\
&= \sum_{ a_\text{trg}, \hat{a}_\text{trg}} \pr{\hat{a}_\text{trg}, a_\text{trg}} \deltap(\hat{a}_\text{trg}, a_\text{trg})
\nonumber \\
&= \sum_{\substack{ a_\text{trg}\\ o_\text{post}\\ \hat{a}_\text{trg}} } \psi(a_\text{trg} | o_\text{pre})
  f(o_\text{post} | a_\text{trg}, o_\text{pre}) h( \hat{a}_\text{trg} | o_\text{pre}, o_\text{post})\nonumber\\
& \qquad \deltap(\hat{a}_\text{trg}, a_\text{trg}).\nonumber
\end{align}

This formula represents the adversary's expected estimation error. As a pessimistic alternative quantification of privacy, one could take the minimum estimation error over all possible values of $a_\text{trg}$, which would correspond to a worst-case scenario.

Regarding $\deltap(\hat{a}_\text{trg}, a_\text{trg})$, intuitively it can be seen as a distance between $\hat{a}_\text{trg}$ and $a_\text{trg}$ that measures the similarity between the inferred and real locations with respect to the user's privacy concerns. For instance, it can be the sum, the minimum, or the maximum of the Euclidean distances between the corresponding locations of each vector, i.e., the total, minimum, or maximum error of the adversary over each pair of corresponding locations in the two vectors. Alternatively, it could be the Hamming distance between the two vectors, i.e., the number of locations at which the estimate differs from the true value.

The function $\deltap(\hat{a}_\text{trg}, a_\text{trg})$ may also be a \emph{weighted} sum, minimum, or maximum to encode the \emph{privacy sensitivity of individual locations} $r\in a_\text{trg}$. For example, when $r$ is a very sensitive location the contribution of estimating $\hat{r} \in \hat{a}_\text{trg}$ to $\deltap(\hat{a}_\text{trg}, a_\text{trg})$ could be large even if $r$ and $\hat{r}$ differ by very little.

Furthermore, $\deltap(\hat{a}_\text{trg}, a_\text{trg})$ can encode the privacy sensitivity of \emph{transitions between locations}, rather than individual locations taken separately. For instance, visiting the bank and visiting a government official may not be very sensitive if considered separately, but visiting the official immediately after the bank may be much more sensitive, especially if the user just made a large withdrawal from the bank and the official is in charge of land development licensing. Moreover, transitions between regions reveal the direction of travel. For instance, the adversary may learn whether the user enters or exits a building, e.g., a hospital.

Depending on the privacy concerns of the user, and whether she wants to protect sensitive locations or sensitive transitions between locations, an appropriate $\deltap(\cdot)$ needs to be chosen. A useful observation is that some $\deltap(\cdot)$ definitions are more general than others: a $\deltap(\cdot)$ that protects the transition between two successive locations automatically protects the locations themselves, so it could be used to protect both objectives simultaneously.

\paragraph*{Quality Metric}

Sending pseudoevents instead of true events to the LBS may help with privacy, but it also degrades the quality of the response that the LBS sends back. We model as follows the quality loss stemming from exposing pseudolocations:

\begin{itemize}
   \item $q_\text{trg}$ denotes the relevant events with respect to quality. Similarly to $a_\text{trg}$, $q_\text{trg}$ is a time-indexed vector. However, its time indices are not necessarily the same as those of $a_\text{trg}$: The locations/times that matter for quality may be different from the ones that matter for privacy.
   \item $\deltaq(q_\text{trg}, o_\text{post}, o_\text{pre})$ represents the quality loss when $q_\text{trg}$ is the true value of the quality-relevant events, the LPPM currently reports $o_\text{post}$ and it has reported $o_\text{pre}$ in the past. The function $\deltaq$ is an input to our framework and it reflects the value of accurate location information for the service provider to return useful service to the user. It also reflect the user's requirements with respect to the quality of the service. Hence, it needs to be determined by the user. 
 \end{itemize}

The expected quality loss caused by an LPPM $f(\cdot)$ is the expected value of $\deltaq(q_\text{trg}, o_\text{post}, o_\text{pre})$ over all $q_\text{trg}$ and $o_\text{post}$, for a given history $o_\text{pre}$:

\begin{align}\label{eq:servicequalityloss}
&Q_{loss}(f, \deltaq, o_\text{pre}) = \ex{\deltaq(q_\text{trg}, o_\text{post}, o_\text{pre})|o_\text{pre}}
\\
&=\sum_{q_\text{trg},o_\text{post}} \pr{q_\text{trg},o_\text{post}|o_\text{pre}}\deltaq(q_\text{trg}, o_\text{post}, o_\text{pre})
\nonumber\\
&=\sum_{q_\text{trg},o_\text{post}} \pr{q_\text{trg}|o_\text{pre}}\pr{o_\text{post} | q_\text{trg}, o_\text{pre}} \deltaq(q_\text{trg}, o_\text{post}, o_\text{pre}) \nonumber
\end{align}

In the equation above, $f(\cdot)$ is hidden in $\pr{o_\text{post} | q_\text{trg}, o_\text{pre}}$, which can be unwrapped as
\begin{align}
&\pr{o_\text{post} | q_\text{trg}, o_\text{pre}} = \sum_{a_\text{trg}} \pr{o_\text{post}, a_\text{trg} | q_\text{trg}, o_\text{pre}}\\
&=\sum_{a_\text{trg}} \pr{o_\text{post}| a_\text{trg}, q_\text{trg}, o_\text{pre}} \pr{a_\text{trg} | q_\text{trg}, o_\text{pre}} \nonumber\\
&=\sum_{a_\text{trg}} f(o_\text{post}| a_\text{trg}, o_\text{pre}) \pr{a_\text{trg} | q_\text{trg}, o_\text{pre}}. \nonumber
\end{align}

We assume that there is a maximum expected quality loss $\dqmax$ that users are willing to tolerate. Formally,
\begin{equation}\label{eq:servicequalityconstraint}
    Q_{loss}(f, \deltaq, o_\text{pre})  \leq \dqmax.
\end{equation}

The quality loss $\deltaq(q_\text{trg}, o_\text{post}, o_\text{pre})$ can be seen as a distance between two vectors: $q_\text{trg}$ and the combination of $o_\text{post}, o_\text{pre}$. It will be zero if an accurate, noiseless trajectory is reported by the LPPM (i.e., if these two vectors are identical), but otherwise it will be positive. If the application needs high location precision to function well, then $\deltaq(\cdot)$ will be large even for a small difference between $q_\text{trg}$ and $o_\text{post}, o_\text{pre}$.

Moreover, $\deltaq(\cdot)$ can encode the variable sensitivity across locations. For example, in a location with many nearby restaurants, an application that finds the nearest restaurant can tolerate a lot of noise, while isolated areas may require more precision. The quality loss function can be instantiated, among other possibilities, as a Euclidean or Hamming distance between real ($q_\text{trg}$) and reported locations ($o_\text{post}, o_\text{pre}$).

The versatility of $\deltaq(\cdot)$ extends to encoding quality loss for applications that depend on the whole \emph{trajectory of locations}, rather than just on a single location. For example, consider a car insurance company that monitors the driving behavior of a customer: quantities such as speed and sudden acceleration or deceleration cannot be evaluated on single locations. Alternatively, one can imagine an discount coupon application that sends different coupons to a user who just visited a sports-shoes store and then a baseball stadium (possibly an amateur baseball player) from a user who visited a general-shoe store after the sports-shoe store (possibly just out buying shoes for the family).

\subsection{Sparse LBS Access Pattern}
We have assumed that the user accesses the LBS at every single time instant although in reality one cannot expect that LBS accesses are uninterrupted (e.g., users may access an online navigation system to travel around some parts of a city, but not in others). In this scenario, the adversary may not only be interested in inferring locations from where the user accesses the LBS, but may also be concerned about the user's whereabouts \emph{between} two LBS accesses, i.e., at times when there are no corresponding exposed pseudolocations. Therefore, these intermediate locations also need to be protected (recall objective (4) in the introduction).

Our dense-LBS-access assumption can also express such a privacy objective. We can accommodate these inference targets by extending the definition of $\deltap(\hat{a}_\text{trg}, a_\text{trg})$ to incorporate the privacy sensitivity of any intermediate locations that the adversary can infer from $\hat{a}_\text{trg}$. For example, assume that, from the estimates $\hat{r}_{t-1}$ and $\hat{r}_t$ of locations $r_{t-1}$ and $r_t$, the adversary can produce estimates for two intermediate locations visited at $t-1+\epsilon$ and $t-\epsilon$ (e.g., the attacker can use Viterbi decoding~\cite{hmm-rabiner} if she wants to infer the most likely trajectory, or the forward-backward algorithm~\cite{hmm-rabiner} if she wants to compute the probability distribution of locations at some times between $t-1$ and $t$). In this case, $\deltap((\hat{r}_{t-1}, \hat{r}_t), (r_{t-1}, r_t))$ would be expressed as $\deltap((\hat{r}_{t-1}, \hat{r}_{t-1+\epsilon}, \ldots, \hat{r}_{t-\epsilon}, \hat{r}_t), (r_{t-1}, r_{t-1+\epsilon}, \ldots, r_{t-\epsilon}, r_t))$. We note that this approach to handling the estimation of such ``in-between'' locations, i.e., locations where the LBS is not accessed, is not particular to LPPMs designed under our framework. It can be used to complement other privacy-preserving solutions in the literature.

\section{Trajectory Privacy as a Stackelberg Game}\label{sec:gameformulation}

\subsection{General Privacy Scenario}

\begin{figure}
  \centering
  \includegraphics[width=\columnwidth, trim=0in 1in 0in 1in, clip]{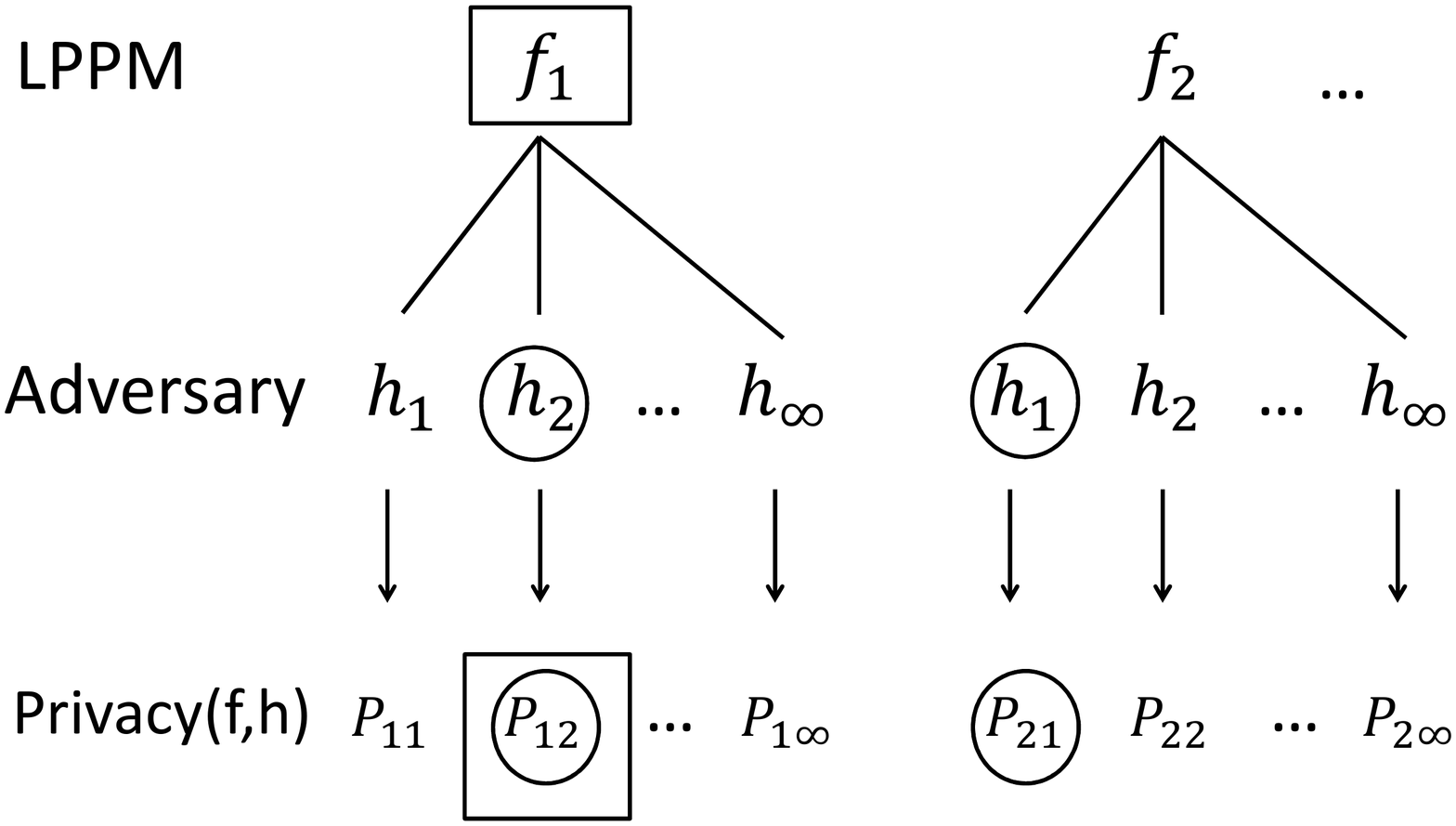}\\
  \caption{The LPPM can choose among an infinite selection of probability functions $f$. For each one of the $f$, the adversary chooses one of his infinite selection of attacks $h$ to minimize Privacy($f,h$); for instance, $h_2$ is the best response to $f_1$, resulting in Privacy $P_{12}$, and $h_1$ is the best response to $f_2$, resulting in Privacy $P_{21}$. Call $h_\text{min}(f)$ the minimizing $h$ for a given $f$. Anticipating the adversary's choice for each $f$, the LPPM chooses the $f$ that maximizes Privacy($f,h_\text{min}(f)$); for instance, if $P_{12} > P_{21}$, the LPPM would choose $f_1$ over $f_2$. Call $f_\text{max}$ the maximizing $f$. The resulting pair $f_\text{max}, h_\text{min}(f_\text{max})$ is the Stackelberg equilibrium, and the privacy achieved is Privacy($f_\text{max}, h_\text{min}(f_\text{max})$).} \label{fig:stackelberg}
\end{figure}

As we have discussed in Section~\ref{sec:statement}, our goal is to design an LPPM that protects user privacy (by maximizing $\deltap$), while preserving quality of service (by respecting the maximum quality loss threshold $\dqmax$ for $\deltaq$). Designing an LPPM reduces to choosing appropriate values for the probabilities $f(o_\text{post} | a_\text{trg}, o_\text{pre})$. In addition, the design process must anticipate that the adversary will know the values chosen for $f$, which means that she will choose the attack $h$ accordingly.

Figure~\ref{fig:stackelberg} details the reasoning involved in solving our task, which is equivalent to solving a Stackelberg game. The distinguishing feature of this game is that there is a \emph{leader}, who commits to a choice, and a \emph{follower}, who observes the leader's choice and then makes a choice of his own. In our task, the leader is the user and her choice is the LPPM, the follower is the adversary who chooses an attack given the user's choice. The Stackelberg equilibrium is a pair of choices ($f^*$ for the LPPM and $h^*$ for the attack) such that neither the user nor the adversary would gain anything by changing their respective choices. In other words, $h^*$ is the choice that minimizes privacy against $f^*$, and $f^*$ is the choice that maximizes privacy against an adversary who will make her choice after observing $f^*$ while respecting the quality constraint. Note that $f^*$ does not necessarily maximize privacy against $h^*$, i.e., if the user could be certain that the adversary would choose $h^*$, then a better choice than $f^*$ could exist. The LPPM design tries to limit the worst possible privacy loss, knowing only that the adversary will choose the most effective attack against whatever $f$ the LPPM implements.

Formally, the general LPPM design task is to choose $f$ and $h$ that solve the Stackelberg game max-minimization
\begin{equation}\label{eq:general-privacy}
  \max_f \min_h Privacy(\psi, f, h, \deltap; o_\text{pre})
\end{equation}
subject to
\begin{equation}\label{eq:general-quality}
    Q_{loss}(f, \deltaq, o_\text{pre}) \leq \dqmax.
\end{equation}

Other than $f$ and $h$, all functions and parameters $\psi$, $\deltap$, $o_\text{pre}$, $\deltaq$, $\dqmax$ are inputs to the problem: $\psi$ -- the user's mobility, and $\deltap$ -- the user's location sensitivity, are specific to the user we aim to protect; the last two $\deltaq, \dqmax$ are specific to the LBS application and perhaps also depend on the user's tolerance to quality deterioration; the prior observations $o_\text{pre}$ depend on the particular time when the user wants to protect her privacy.


We now give two specific examples of the general LPPM design task for two plausible privacy objectives. The first objective is to protect the $k+1$ most recent locations (including the current one at time $t$), having already exposed pseudolocations for the $k$ past time instants, by choosing an appropriate pseudolocation to expose at time $t$. The second objective is to protect the current \emph{and} future locations, assuming nothing has been exposed so far, by choosing a pseudolocation for time $t$ \emph{and} for future time instants. We emphasize that one can instantiate any number of objectives by selecting the time period one wishes to protect ($a_\text{trg}$), the events that have been exposed already ($o_\text{pre}$), and the time period for which the LPPM can expose pseudolocations ($o_\text{post}$).

\subsection{Joint Protection of Past-Present Locations}\label{subsec:pastpresent}

Consider a user who, at time $t$, wants to publish her location $r_t$. She has already published her locations $r_\tau$ at the $k$ previous time instants $\tau = \{t-1, t-2, \ldots, t-k\}$. These locations have been sent by the LPPM to the LBS as pseudolocations $o_\tau$. As explained in Section~\ref{sec:statement}, locations exposed prior to time $t-k$, i.e., from $1$ to $t-k-1$, are considered to have no influence on the choice of the user at time $t$.

The simplest privacy objective ($a_\text{trg}$) that the user could have is to protect her current location only: $a_\text{trg} = r_t$. We call this \emph{single location privacy}. But as we have argued in \iflongver Section~\ref{sec:lbs}, \fi the \emph{transition} from the previous location $r_{t-1}$ to the current position may be sensitive, or, in general, the transition from the $k$-tuple $r_\tau$ to $r_t$ may be sensitive. A different objective, therefore, is to protect the whole vector $r_\tau$ in addition to $r_t$: $a_\text{trg} = (r_t, r_\tau)$. Observe that the latter objective (transition privacy) is more general the previous one (single location privacy): Choosing a function $\deltap((\hat{r}_t,\hat{r}_\tau), (r_t,r_\tau))$ that just ignores $r_\tau$ and $\hat{r}_\tau$ makes the two cases equivalent.

The prior observations $o_\text{pre}$ are the set of pseudolocations $o_\tau$, and the pseudolocation $o_\text{post}$ that the LPPM produces is just the one corresponding to the current time $t$: $o_t$. The quality loss $\deltaq$ is a function of past and present exposed pseudolocations, $(o_t, o_\tau)$, and of $q_\text{trg}$. As stated in Section~\ref{sec:statement} $q_\text{trg}$ does not need to coincide with $a_{trg}$, and can consist of any subset of events from time 1 up to and including $t$. In fact, it does not even need to overlap with $(r_t, r_\tau)$.



%
%
%
%
%
%
%
%
%

Making the appropriate substitutions in \eqref{eq:privacygain}, we derive the privacy definition for the specific case of protecting past and present locations as follows:
\begin{equation}\label{eq:pastpresentprivacy}
\begin{split}
  Privacy(\psi, f, h, \deltap; o_\tau) = \sum_{r_t, r_\tau, o_t, \hat{r}_t,\hat{r}_\tau}\psi(r_t, r_\tau | o_\tau)\\
  f(o_t | r_t, r_\tau, o_\tau)\\
  h(\hat{r}_t, \hat{r}_\tau|o_t, o_\tau)\\
  \deltap((\hat{r}_t,\hat{r}_\tau), (r_t,r_\tau))
\end{split}
\end{equation}
The quality loss is also straightforward to define using \eqref{eq:servicequalityloss}:
\begin{equation}\label{eq:pastpresentquality}
\begin{split}
Q_{loss}(f, \deltaq, o_\text{pre})& = \ex{\deltaq(q_\text{trg}, o_t, o_\tau)|o_\tau}\\
& =\sum_{q_\text{trg}, o_t} \pr{q_\text{trg}, o_t|o_\tau} \deltaq(q_\text{trg}, o_t, o_\tau)\\
& =\sum_{q_\text{trg}, o_t} \pr{q_\text{trg} |o_\tau} \pr{o_t | q_\text{trg}, o_\tau} \deltaq(q_\text{trg}, o_t, o_\tau)
\end{split}
\end{equation}


Notice that setting $k=0$ eliminates $o_\text{pre}$ (which would mean that none of the previous pseudolocations exposed to the adversary are assumed to correlate with the current location), and the target events $a_\text{trg}$ reduce to the current time $t$ only. In total, only the current location matters for privacy and for quality, and the design task reduces to the sporadic case handled by Shokri~et~al.'s framework~\cite{ShokriTDHL11}.

\subsection{Joint Protection of Present-Future Locations}\label{subsec:presentfuture}

We now consider a user who, as before, wants to publish her location $r_t$ at time $t$. However, in contrast with the previous case, she is not concerned about the past locations she has visited, but rather about future ones. This concern can be motivated as follows:

Disclosing the current location might not be important in and of itself, but it might make it much easier for the adversary to infer the next location, which happens to be very sensitive. For instance, the user might currently be on a street that only leads to an abortion clinic. Hence, disclosing her current location is almost equivalent to disclosing that she will go to the clinic. Symmetrically, her current location might be very sensitive, and her next (expected) location can be linked easily to her current one. For instance, she might about to leave the abortion clinic and enter a street that is only used as the clinic's exit. Furthermore, as argued in \iflongver Section~\ref{sec:lbs}, \fi neither the current nor the next location might be particularly sensitive separately, but the transition from one to the other might be.

The conclusion in all these cases is that the current location must be protected jointly with the (possible) next one(s), where the user will be at time $t+1$ and later. For this reason, $a_\text{trg}$ includes time subscripts larger than $t$, and so does $o_\text{post}$, i.e., the LPPM should take into account at the present time $t$ what it is likely to output in future times, so that the current choice of $o_t$ does not limit future choices. The intuition is that the LPPM should choose the current pseudolocation $o_t$ so that future paths that the user will likely take can be protected with pseudolocations that are compatible with $o_t$.




For simplicity, we consider an example where (a) the LPPM anticipates only the next location, rather than many successive future locations, so $a_\text{trg}$ is $(r_{t+1}, r_t)$ and $o_\text{post}$ is $(o_{t+1}, o_t)$, and (b) no prior locations matter for privacy, so $ o_\text{pre}$ is omitted.


Substituting in \eqref{eq:privacygain}, privacy in this example is defined as
\begin{multline}
  Privacy(\psi, f, h, \deltap) = \\
  \sum_{\substack{r_{t+1}, r_t\\  o_{t+1}, o_t \\ \hat{r}_{t+1}, \hat{r}_t} } \psi(r_{t+1}, r_t)
  f( o_{t+1}, o_t| r_{t+1}, r_t )\\
  h( \hat{r}_{t+1}, \hat{r}_t | o_{t+1}, o_t )
  \deltap( (\hat{r}_{t+1}, \hat{r}_t), (r_{t+1}, r_t) ),
\end{multline}
and the quality loss, from \eqref{eq:servicequalityloss}, is defined for a general $q_\text{trg}$ as
\begin{multline}
Q_{loss}(f, \deltaq) = \ex{\deltaq(q_\text{trg}, o_t, o_{t+1})}\\
=\sum_{\substack{q_\text{trg}, o_t, o_{t+1}} } \pr{q_\text{trg}, o_t, o_{t+1}}  \deltaq(q_\text{trg}, o_t, o_{t+1}).
\nonumber 
\end{multline}


\subsection{Optimal Attacks and Defenses via Linear Programming}\label{sec:solution}

Having reduced the LPPM design to a Stackelberg game max-minimization, we now compute the equilibrium of the game, which is equivalent to computing the optimal defense $f$ and attack $h$.

Note that there is an infinity of candidate $f$s and $h$s (all possible probability distributions), so enumeration (as shown in Figure~\ref{fig:stackelberg}) cannot be used directly as an algorithm to find the equilibrium. To overcome this problem, we use a standard technique for transforming the computation of a game theoretic equilibrium to a linear program. This technique introduces auxiliary variables $x_{o_\text{post}}$, for each possible value of $o_\text{post}$, which roughly correspond to the amount of privacy gained when the LPPM reports each particular value of $o_\text{post}$. As previous research \cite{ShokriTTHB12} has expanded on this transformation technique, we merely present the key features of the resulting linear program:




We want to maximize $\sum_{o_\text{post}} x_{o_\text{post}}$ under the constraint
\begin{multline}  \label{eq:constrPriv}
  x_{o_\text{post}} \leq \sum_{a_\text{trg}} \psi(a_\text{trg} | o_\text{pre}) f(o_\text{post} | a_\text{trg}, o_\text{pre}) \deltap(\hat{a}_\text{trg}, a_\text{trg}),  \\   \forall \hat{a}_\text{trg}, o_\text{post},
\end{multline}
and under the constraint
\begin{multline}\label{eq:quality}
  \sum_{\substack{q_\text{trg}\\ o_\text{post}}} \pr{q_\text{trg}|o_\text{pre}} \pr{o_\text{post} | q_\text{trg}, o_\text{pre}} \deltaq(q_\text{trg}, o_\text{post}, o_\text{pre}) \\ \leq \dqmax.
\end{multline}
Equation~\eqref{eq:constrPriv} is equivalent to the min-maximization \eqref{eq:general-privacy}, and equation~\eqref{eq:quality} is just the quality constraint \eqref{eq:general-quality}.

In this way, we can compute the optimal LPPM $f^*$. The optimal attack $h^*$ can either be computed via \eqref{eq:general-privacy}, or by solving another linear program called the dual. We do not go into detail about the dual here and point the interested reader to prior research \cite{ShokriTTHB12} and to standard textbooks on linear programming \cite{vanderbei2008linear}.

Having computed the optimal LPPM $f^*$ and the optimal attack $h^*$, the resulting $Privacy(f^*, h^*)$ is the level of privacy achieved by the user, and the resulting $Q_{loss}(f^*)$ is the application's incurred quality loss.

\noindent\textbf{Alternative quantifications of privacy.  }
As we mentioned at the beginning of this section, the optimal LPPM $f^*$ depends on all inputs $\psi, \deltap, o_\text{pre}, \deltaq, \dqmax$, so if any of these functions or variables changes, a new LPPM must be computed to maximize privacy. Alternatively, it is possible to compute an LPPM that maximizes the \emph{average} privacy across a range of different values, e.g., $o_\text{pre}$, or across a range of different applications represented by different functions $\deltaq$. One could also be conservative and compute instead an LPPM that maximizes the \emph{minimum} privacy across a range of values, e.g., $o_\text{pre}$. In this paper, we take $\psi, \deltap, o_\text{pre}, \deltaq, \dqmax$ to be given inputs, but we note that it is not difficult to adapt the framework to accommodate alternatives.

\section{Evaluation}\label{sec:evaluation}

By formulating the LPPM design as an optimization problem, the LPPM algorithm $f(\cdot)$ is guaranteed to be optimal among all possible algorithms that respect the same constraints. Hence, there is no point in evaluating our design with simulations or any other heuristic evaluation method. We nevertheless compare to a sporadic LPPM to stress the importance of using a trajectory-aware LPPM. The sporadic LPPM that we use is the optimal one, as presented in prior work~\cite{ShokriTTHB12}. Due to incompatible assumptions, we cannot compare to the trajectory-aware LPPMs in the literature (see discussion of related work in Section~\ref{sec:relatedwork}). We also show, for two illustrative scenarios, how our LPPM design allows to trade off privacy and quality. Finally, we discuss the run-time complexity of our design.\iflongver\footnote{The optimization problem involved in the LPPM design (see Section~\ref{sec:solution}) can be solved with any linear programming software. In our evaluation, we used MATLAB's \texttt{linprog()} function.}\fi

For the comparison to the sporadic LPPM and for the illustration of the privacy-quality tradeoff, we use a real data set of location traces. These traces, which are one day long, belong to $10$ randomly chosen mobile users (vehicles) in the San Francisco Bay area from the epfl/mobility dataset at CRAWDAD \cite{CRAWDAD}. These $10$ examples serve to illustrate the optimality of the LPPMs designed by our method, since the technique is user-centric and does not need information about other users.

We discretize both time and location: we divide the Bay Area into $10\times25$ equal-size locations, and consider a day to be composed by $288$ time units, one per each $5$ minutes. We emphasize that the granularity of both time and locations can be arbitrarily selected depending on the required accuracy in quantifying privacy and service quality.\footnote{Note that locations need not necessarily form a grid.
In general, a higher number of locations corresponds to higher granularity, and, as a result, to more precise and more accurate quantification and protection of location privacy. The only requirement is for time and locations to be discrete. Of course, more locations and more time instants make the computation of the LPPM more demanding in resources. The run-time computation discussion in \iflongver Section~\ref{sec:computation} \fi elaborates on the effect of the number of time instants and locations on the complexity of the problem.} We consider all the locations that are visited by each user, which on average is $23.4$ locations per user. We also consider all the transitions that each user has made between these locations in our dataset.



For both the comparison to the sporadic LPPM and for the privacy-quality tradeoff, we need to specify all the input parameters/functions $\deltap$, $\deltaq$, $\dqmax$, $o_\text{pre}$, $\psi$.

Without loss of generality, we select the privacy gain $\deltap$ and the quality loss $\deltaq$ functions to be the Hamming distance: $\deltap(\hat{a}_{trg}, a_{trg}) = 1_{\hat{a}_{trg} \neq a_{trg}}$ and $\deltaq(q_\text{trg}, o_\text{post}, o_\text{pre}) = 1_{q_\text{trg} \neq (o_\text{post}, o_\text{pre})}$. Using the Hamming distance means, taking the privacy gain as an example, that the only bad case for privacy is when the attacker correctly estimates the \emph{exact} value of the target locations (i.e., when $\hat{a}_{trg}$ is exactly equal to $a_{trg}$). All other estimates are equally good for privacy, regardless, e.g., of the physical distance between the attacker's estimate and the true value of $a_{trg}$. As our quantification of privacy is the \emph{expected} value of $\deltap(\hat{a}_{trg}, a_{trg})$ -- and the expected value of $1_{\hat{a}_{trg} \neq a_{trg}}$ is just the probability of $\hat{a}_{trg} \neq a_{trg}$ -- in effect we quantify privacy as the probability that the adversary will make an erroneous estimate.

For the maximum tolerable quality loss $\dqmax$, we do not specify a single value, but rather compute the achievable privacy for multiple values, so as to observe the privacy-quality tradeoff.

For the previously reported events $o_\text{pre}$, we do not specify a single value. Instead, the privacy values that we compute and present in the following figures are averaged over all possible values of $o_\text{pre}$, because such an average is more representative of the privacy that a user can expect to achieve: $$\sum_{o_\text{pre}} \pr{o_\text{pre}} Privacy(\psi, f, h, \deltap; o_\text{pre}).$$
We must stress that the actual values obtained in the evaluation depend on context (e.g., user mobility). In this sense, they amount to the ``total'' privacy of the user, not just the privacy that is due to the LPPM alone. However, the evaluation shows that, other context being the same, our LPPMs achieve the highest ``total'' privacy value among all mechanisms.

To compute the prior probability $\psi(a_\text{trg})$ on the target events, we use the aforementioned traces to build a first-order Markov chain on the discretized set of locations. Choosing a Markov chain over a different mobility model is arbitrary. As mentioned in Section~\ref{sec:statement}, our method can handle other mobility models as well.

In relation to $\psi$, notice that, in general, we need to specify the \emph{conditional} prior on the target events $\psi(a_\text{trg} | o_\text{pre})$, i.e., the prior \emph{given} the previously reported events $o_\text{pre}$ (see Equation~\eqref{eq:privacygain}). But from the traces we can only compute the \emph{unconditional} prior $\psi(a_\text{trg})$. The connection between $o_\text{pre}$ and $a_\text{trg}$ will typically be given by whatever LPPM was in use when $o_\text{pre}$ was reported, in conjunction with the unconditional prior. In the Appendix, we show how this connection can be established, and how $\psi(a_\text{trg} | o_\text{pre})$ can be computed, for the joint protection of past and present privacy that we illustrate in this section.

\begin{figure}[t]
    \centering
    \includegraphics[width=\columnwidth]{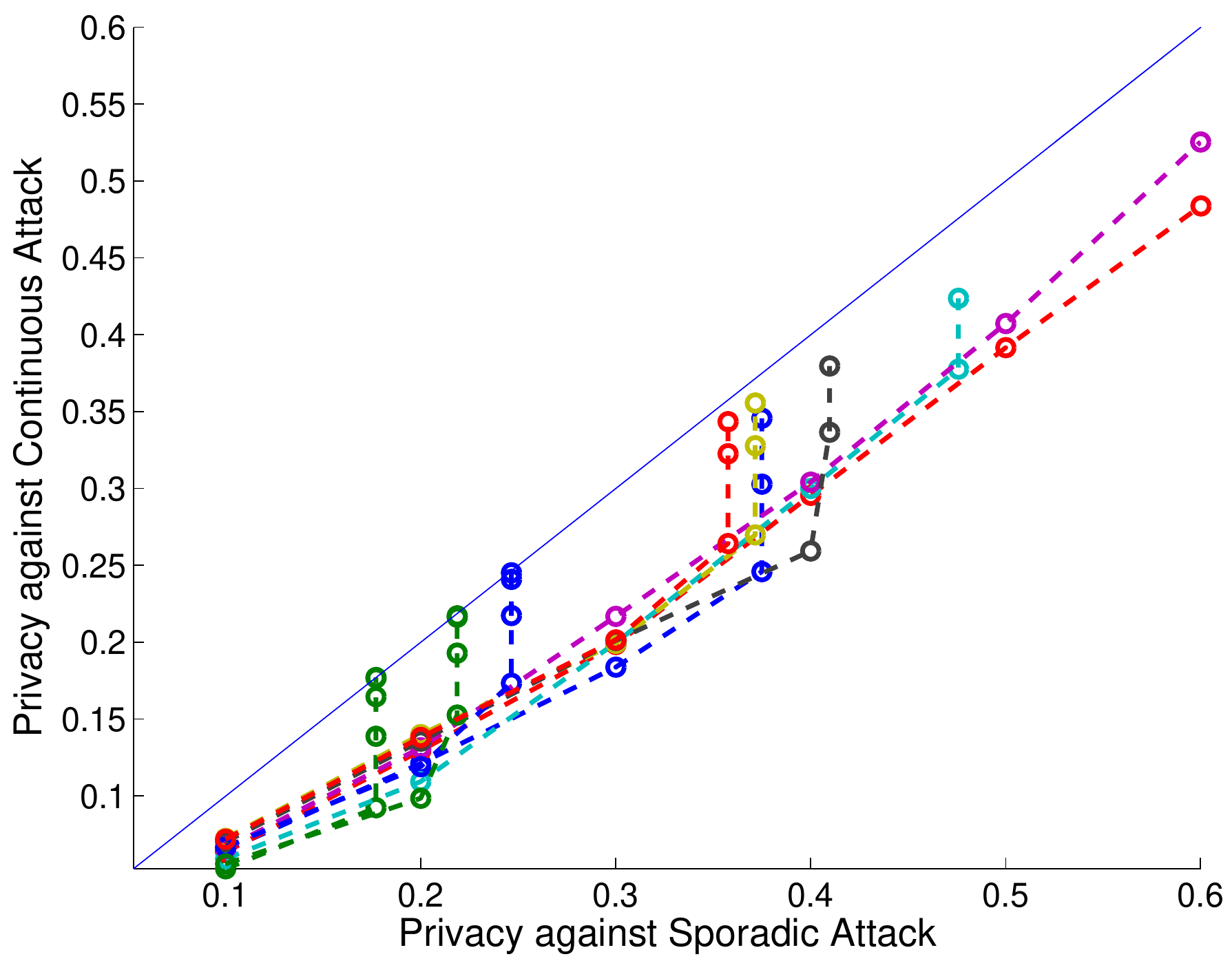}
    \caption{Users' single-location privacy, using a sporadic LPPM against two attacks: sporadic attack vs. correlation-aware attack. For 10 different users (lines), and for various values of the service quality threshold $\dqmax$ (dots), we see that the privacy against a correlation-aware attack (vertical axis) is always less than the privacy against a sporadic attack (horizontal axis).}
    \label{fig:privacy:single:sporadic-vs-continuous}
\end{figure}

\begin{figure}[h]
    \centering
    \includegraphics[width=\columnwidth]{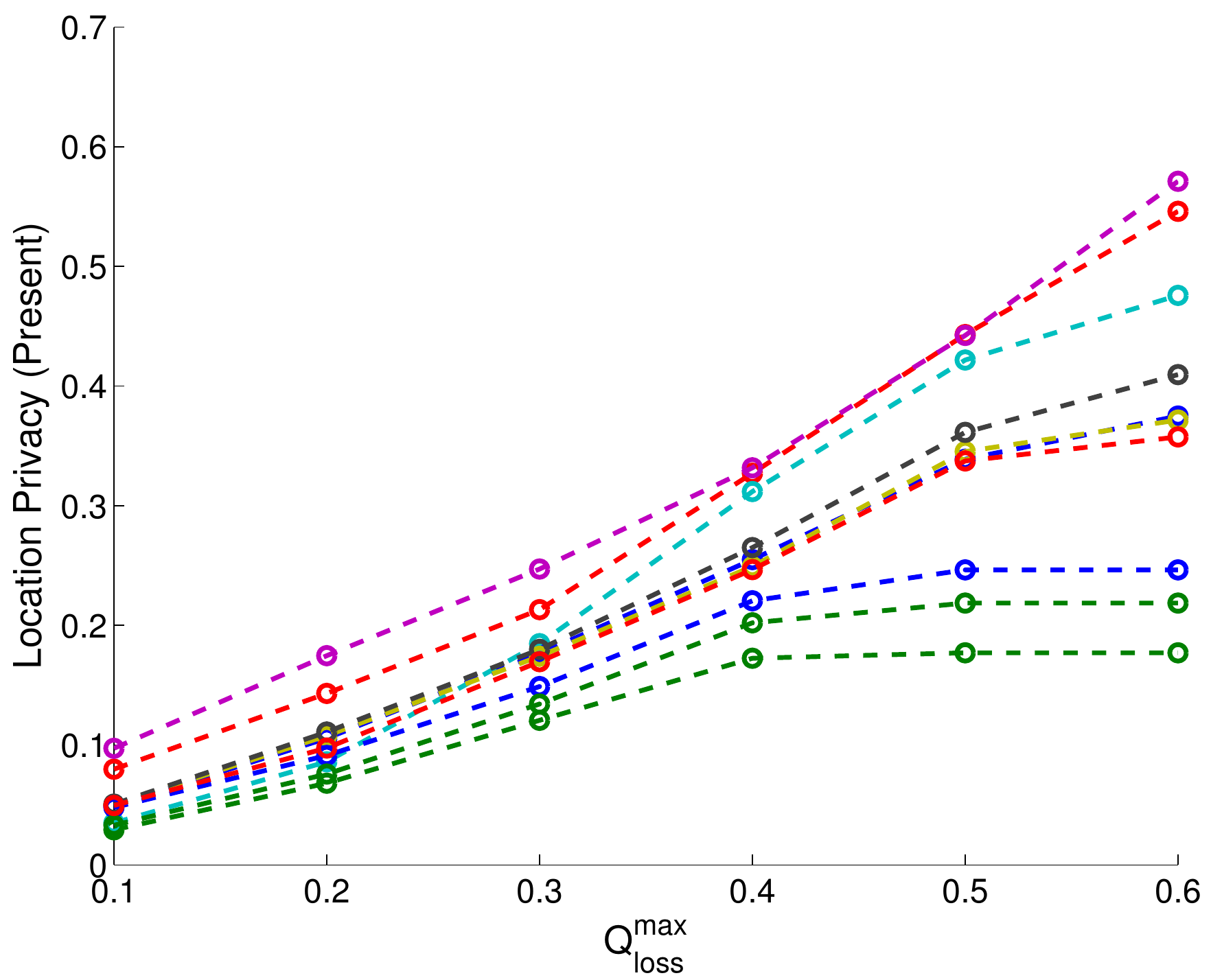}
    \caption{Privacy-quality tradeoff in the first scenario (single-location privacy): Protecting the current location ($a_\text{trg} = r_t$), when $o_\text{pre}$ is the pseudolocation reported in the previous time instant $o_{t-1}$. Each curve corresponds to one user.}
    \label{fig:privacy:single:timethree}
\end{figure}

\begin{figure}
    \centering
    \includegraphics[width=\columnwidth]{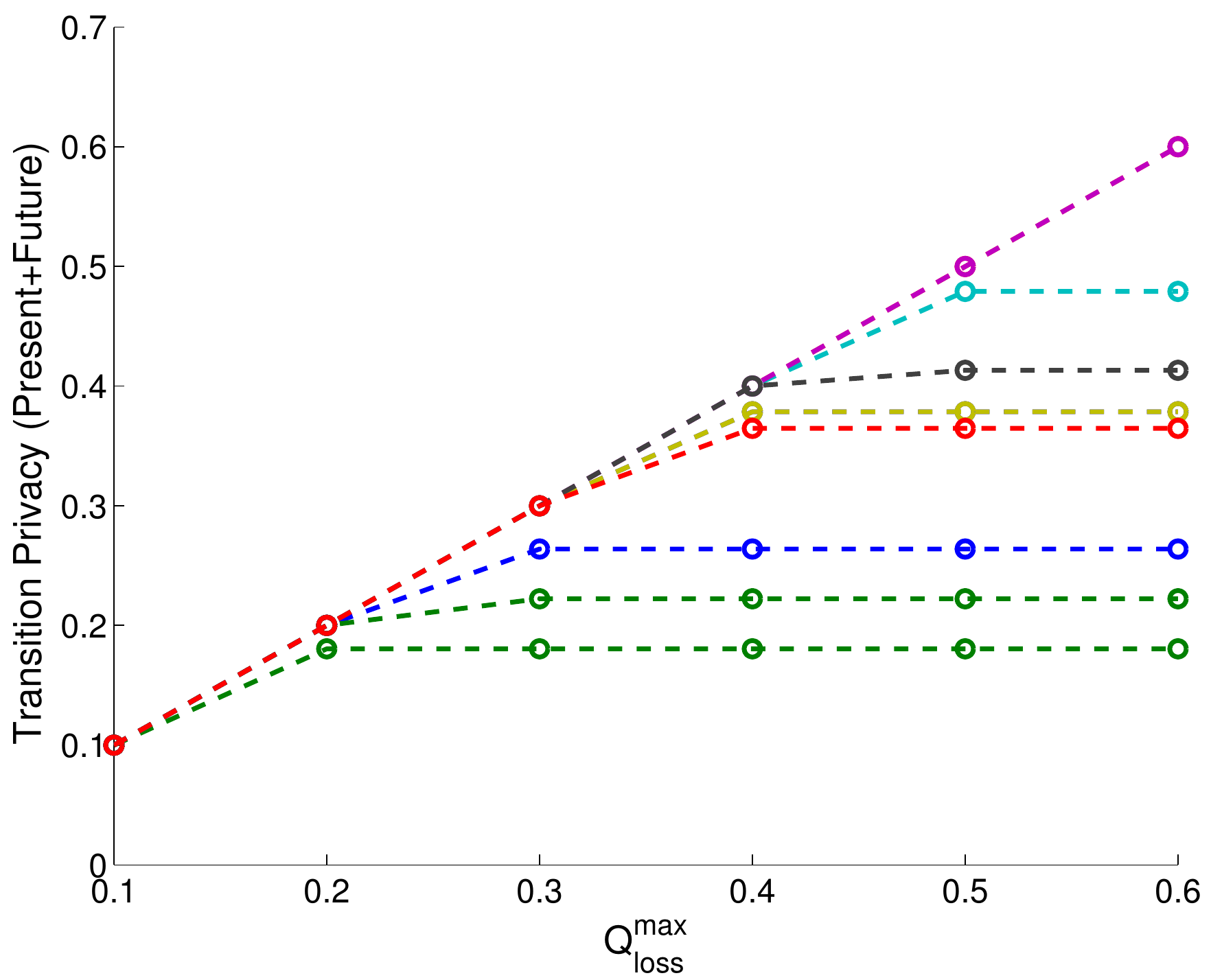}
    \caption{Privacy-quality tradeoff in the second scenario (transition privacy): Protecting the current and the next location ($a_\text{trg} = (r_t, r_{t+1})$). Each curve corresponds to one user.}
    \label{fig:privacy:transition:future}
\end{figure}

\subsection{Comparison to Optimal Sporadic LPPM}

A trajectory-oblivious (sporadic) LPPM is typically evaluated against an attack that is also sporadic, i.e., an attack in which location correlation is not taken into account. To provide quantitative justification for the inadequacy of such LPPMs and their evaluation when the exposed locations are correlated, we show in Figure~\ref{fig:privacy:single:sporadic-vs-continuous} that a correlation-aware attack can achieve much lower privacy than a sporadic attack.

Of course, a sporadic LPPM protects single locations only, so to compare meaningfully, we pick as objective of the correlation-aware attack the single-location privacy objective, i.e., $a_\text{trg} = r_t$ (see Section~\ref{subsec:pastpresent}). The difference between the correlation-aware attack and the sporadic attack is that the former uses the conditional prior probability on the target location $\psi(r_t | o_\text{pre})$ (for $o_\text{pre} = o_{t-1}$), whereas the latter uses the unconditional prior $\psi(r_t)$.

Each attack is paired against the same sporadic LPPM (the optimal one \cite{ShokriTTHB12}, as mentioned earlier), and the results are plotted across the 10 mobile users and for various values of the service quality threshold $\dqmax$. As all data points are below the $x=y$ diagonal, we conclude that privacy in the correlation-aware attack ($x$-axis) is lower than privacy in the sporadic attack ($y$-axis). The only cases where the two attacks are equally (un-)successful are when the quality loss threshold is so high that the sporadic LPPM can inject enough noise to blur even the inference of a correlation-aware attack.

\subsection{Privacy-Quality Tradeoff}

In this section, we illustrate the privacy-quality tradeoff of our LPPMs for two particular scenarios: Protecting single-location privacy for the current location, taking into account the immediately previous pseudolocation ($a_\text{trg} = r_t$ and $o_\text{pre} = o_{t-1}$), shown in Figure~\ref{fig:privacy:single:timethree}; and protecting transition privacy for the current and future locations ($a_\text{trg} = (r_t, r_{t+1})$), as described in Section~\ref{subsec:presentfuture}, shown in Figure~\ref{fig:privacy:transition:future}.

Under each of these two scenarios, we construct the optimal protection mechanism for each of the 10 users in our traces (i.e., the mechanism that provides the maximum privacy for her). We plot this maximum privacy as a function of the service quality threshold $\dqmax$. We see in both figures that the achievable privacy increases as $\dqmax$ increases. This is not surprising, as higher values of $\dqmax$ let the LPPM inject more and more noise.

However, in both scenarios we observe two effects: First, a saturation effect takes place for most users as $\dqmax$ increases. Their privacy reaches a plateau beyond which any further increase in $\dqmax$ does not contribute to a corresponding increase in privacy. Second, the privacy plateau, as well as the privacy level for any value of $\dqmax$, is not the same for all users. This suggests that not all users can be protected equally well, but rather there is some inherent privacy limit for each user that is connected to the user's mobility. Users with more predictable mobility cannot be protected as efficiently as less predictable ones, regardless of the amount of noise that the LPPM injects. Looking at the figures, more predictable users correspond to the lowest curves in Figures~\ref{fig:privacy:single:timethree}~and~\ref{fig:privacy:transition:future}, whereas users with more random mobility correspond to the highest curves.

It is worth noting at this point that both these effects are not artefacts of our LPPM. Our LPPMs provide the best possible protection, so these effects are inherent in the mobility patterns of the users.

\iflongver
    \subsection{Computational Considerations} \label{sec:computation}

Our mechanism is intended to be computed offline and used online: The LPPM function $f(o_\text{post} | a_\text{trg}, o_\text{pre})$ is precomputed offline and then downloaded to the device. Then, whenever the user attempts to expose a location, the LPPM looks up and performs the appropriate randomization on pseudolocations $o_\text{post}$, based on the actual values of the target events to be protected $a_\text{trg}$ and the previously exposed vector of pseudolocations $o_\text{pre}$. In this way, the only computational burden of the resource-constrained mobile device is a look-up and a randomized selection of $o_\text{post}$.

The offline computation of the LPPM function $f$ requires solving a separate linear program for each value of $o_\text{pre}$ that may arise in practice. But most of the theoretically possible values of the vector $o_\text{pre}$ are nonsensical sequences of locations, e.g., sequences where successive locations are too far away from each other, so these need not be taken into account, which saves considerable time. Similarly, the number of variables in each linear program is theoretically equal to the total number of pairs of $a_\text{trg}$ and $o_\text{post}$ vectors, since a value for $f$ must be computed for each such combination. This number is $M^{\text{length}(a_\text{trg}) + \text{length}(o_\text{post})}$ (recall that $M$ is the total number of locations -- see Section~\ref{sec:statement}), but in practice it is much smaller. The actual number of linear programs and of variables is closer to the number of likely trajectories of the corresponding length (the number of linear programs is equal to the number of trajectories of length $\text{length}(o_\text{pre})$, whereas the number of variables is equal to the number of trajectories of length $\text{length}(a_\text{trg}) + \text{length}(o_\text{post})$).

It is very important to notice also that the computation of $f$ needs to be done only once, so the associated cost only needs to be incurred once. A recomputation of $f$ is only necessary if, for example, the user parameters or application parameters $\deltap, \deltaq, \dqmax$ change, or if the user wants to protect a different aspect of her privacy (e.g., previous, present, and next location, instead of just present and next location), which would translate to a change in $a_\text{trg}$, or if one wishes to take into account different prior knowledge of previously reported pseudolocations $o_\text{pre}$ (e.g., take into account the 3 previously reported pseudolocations instead of just one).

\fi

\section{Conclusions} \label{sec:conclusion}

Existing location privacy-preserving mechanisms either ignore the information leaked by the exposure of correlated locations, or ignore that the adversary will adapt his attack to the protection mechanism. Hence, in practice, these schemes do not provide the promised level of privacy. In this paper, we have proposed a framework that simultaneously considers correlation and the background knowledge of the adversary, namely the mobility profile of the user, the previously exposed locations, and the internal algorithm implemented by the protected mechanism; while at the same time respecting the user's service quality requirements.

Our framework allows users to design LPPMs that protect not only her current location, but also her past and future whereabouts. Furthermore, our solution is the first to deal with protecting the privacy of transitions between locations, and with preserving the privacy of locations from which the user does not access the location based service. Two key advantages of the framework are that it is not limited to a particular scenario, but can be used to compute optimal defenses for different privacy and quality user preferences; and that it finds an optimal defense among a wide variety of conceivable mechanisms, effectively any mechanism that can be modeled as a probability distribution describing how obfuscated locations are produced from real locations.

Using real mobility traces, we show that users can relax their quality of service requirements in exchange for privacy, but the predictability of their movements determines the maximum protection they can obtain. The privacy level achieved by the LPPMs computed using our framework can be considered an upper bound on the privacy achievable by any defense in presence of a strategic adversary who knows the users' mobility patterns. Hence, our solution is ideal to be used as benchmark to measure the effectiveness of future defenses.

\section*{Acknowledgments}

Carmela Troncoso's research is partially supported by EU 7th Framework Programme (FP7/2007-2013) under grant agreements 610613 (PRIPARE) and 285901 (LIFTGATE). George Theodorakopoulos's research was partially supported by EU FP7 EINS (grant agreement No 288201).

\bibliographystyle{abbrv}
\bibliography{references}

\iflongver
    \appendix
    \section*{Computing the Conditional Prior}


As mentioned in Section~\ref{sec:evaluation}, function $\psi(r_t, r_\tau | o_\tau)$ needs to be specified as input to the linear program \eqref{eq:pastpresentprivacy}, which computes the LPPM for the joint protection of past and present. We now describe how this computation can be performed for the case $k=1$, i.e., when $o_\text{pre} = o_\tau = o_{t-1}$, and the objective is to protect only the current and the previous locations: $a_\text{trg} = (r_t, r_\tau) = (r_t, r_{t-1})$.

In general, recall that $t$ takes values from $\set{T} = \{1, 2, ...\}$. We first describe the cases $t=1$ and $t=2$, which form the base cases of the recursion, and then we handle the general case $t>2$.

If $t=1$, variables $r_{t-1}$ and $o_{t-1}$ do not make sense, as there is no previous LBS-access location nor observation. In this case, $\psi(r_t, r_{t-1} | o_{t-1})$ is just $\psi(r_t)$ and it is equal to the steady state probability $\psi(r), r \in \set{R}$ as computed from the transition matrix $P$. Then, the solution $f$ of the linear program is equivalent to a sporadic LPPM~\cite{ShokriTTHB12}, so we call it $f_\text{spor}$.

If $t=2$, the computation proceeds as follows:
\begin{equation}
  \psi(r_t, r_{t-1} | o_{t-1}) = \pr{r_t | r_{t-1}, o_{t-1}}\pr{r_{t-1} | o_{t-1}}
\end{equation}
But $ \pr{r_t | r_{t-1}, o_{t-1}} = \pr{r_t | r_{t-1}}$, which is known from the transition matrix $P$. So we only need to compute $\pr{r_{t-1} | o_{t-1}}$, which we do with Bayes' rule:
\begin{equation}
  \pr{r_{t-1} | o_{t-1}} = \frac{ \pr{o_{t-1} | r_{t-1}} \pr{r_{t-1}} }
                                { \sum_{r_{t-1}} \pr{o_{t-1} | r_{t-1}} \pr{r_{t-1}} }
\end{equation}
Now, $\pr{r_{t-1}}$ is known (it is the steady state of $P$), so we only need to compute $\pr{o_{t-1} | r_{t-1}}$.

Only in the case of $t=2$, it holds that $\pr{o_{t-1} | r_{t-1}} = f_\text{spor}(o_{t-1} | r_{t-1})$, and $f_\text{spor}$ is computed in the $t=1$ step. This concludes the case $t=2$.

In the general case, $t>2$, we derive $\psi(r_t, r_{t-1} | o_{t-1})$ just as for $t=2$ up to the application of Bayes' rule. The difference is that we can no longer substitute $f_\text{spor}$ for $\pr{o_{t-1} | r_{t-1}}$, so we need to compute it directly:
\begin{multline}
  \pr{o_{t-1} | r_{t-1}} = \sum_{o_{t-2}, r_{t-2}}   \pr{o_{t-1}, r_{t-2}, o_{t-2} | r_{t-1}}\\
                         = \sum_{o_{t-2}, r_{t-2}}   \pr{o_{t-1}| r_{t-1}, r_{t-2}, o_{t-2}}\pr{r_{t-2}, o_{t-2} | r_{t-1}}\\
                         = \sum_{o_{t-2}, r_{t-2}}   \pr{o_{t-1}| r_{t-1}, r_{t-2}, o_{t-2}}\pr{o_{t-2} | r_{t-2}, r_{t-1}} \\
                         \pr{r_{t-2} | r_{t-1}}
\end{multline}
The first term is $f$ as computed for time $t-1$. The third term is known from the transition matrix $P$. The second term $\pr{o_{t-2} | r_{t-2}, r_{t-1}}$ is equal to $\pr{o_{t-2} | r_{t-2}}$, because the obfuscation at time $t-2$ depends only on $r_{t-2}, r_{t-3}$, and $o_{t-3}$. Knowing $r_{t-1}$ when $r_{t-2}$ is already known gives us no extra information on $r_{t-3}$ or $o_{t-3}$. Hence, the computation of $\pr{o_{t-1} | r_{t-1}}$ is shown to be recursive:
\begin{multline}
\pr{o_{t-1} | r_{t-1}} = \sum_{o_{t-2}, r_{t-2}}   f(o_{t-1}| r_{t-1}, r_{t-2}, o_{t-2})\\
\pr{o_{t-2} | r_{t-2}} \pr{r_{t-2} | r_{t-1}}.
\end{multline}

After computing $\psi(r_t, r_{t-1} | o_{t-1})$, we can solve the linear program and find the optimal LPPM for the user, which is dependent on the previous observed location of the user.


\fi
\end{document}